\newcommand{\thj}[6]{
                 \left(\begin{array}{ccc}#1&#2&#3\\#4&#5&#6\end{array}\right)
                    }

\documentstyle[prd,aps,eqsecnum]{revtex}

\begin{document}

\title{Fluctuations of the vacuum energy density of quantum fields in
curved spacetime 
via generalized zeta functions}
\author{Nicholas G. Phillips  and B. L. Hu\\
{\small Department of Physics, University of Maryland, College Park, 
         Maryland 20742} }
\date{UMDPP\#97-58, submitted to Phys. Rev. D, 4 November, 1996}
\maketitle
\begin{abstract}
For quantum fields on a curved spacetime with an Euclidean section,
we derive a general expression for the stress energy tensor two-point
function in terms of the effective action.
The renormalized two-point function is given in terms of the
second variation of the Mellin transform of the trace of the heat kernel
for the quantum fields.
For systems for which a spectral decomposition of the wave opearator is possible,
we give an exact expression for this two-point function.
Explicit examples of the variance to the mean ratio
$\Delta' = \frac{\langle\rho^2\rangle-\langle\rho\rangle^2}
{\langle\rho\rangle^2}$ of the vacuum energy density $\rho$
of a massless scalar field are computed for the spatial 
topologies of $R^d\times S^1$ and $S^3$, with results of
$\Delta'(R^d\times S^1) =(d+1)(d+2)/2$, 
and $\Delta'(S^3) = 111$ respectively.
The large variance signifies the importance of quantum fluctuations and
has important implications for the validity of semiclassical gravity theories
at sub-Planckian scales.
The method presented here can facilitate the calculation of stress-energy
fluctuations for quantum fields useful for the analysis of
fluctuation effects and critical phenomena in problems ranging from atom optics
and mesoscopic physics to early universe and black hole physics.
\end{abstract}

\section{Introduction}

Many important physical processes involve vacuum fluctuations of quantum fields.
Famous examples are Casimir effect \cite{Casimir} and Lamb shift.
These effects shed light on some basic issues of quantum field theory, and
guide the beginning probes and queries into establishing a viable theory
of quantum fields. In curved or topologically non-trivial spacetimes
(see, e.g., \cite{BirDav}),
basic issues such as the ambiguity in the definition of vacuum states, particles, energy,
and the regularization of the energy-momentum tensor occupied the
central attention of investigators in the 70's. Amongst the many formalisms
developed, the zeta-function \cite{DowCri,Haw} and point-separation
\cite{DeW,Chr} regularizations are particularly relevant to our problem.
In the 80's, two directions were noteworthy in this field:
the backreaction of particle creation on the dynamics of the 
universe \cite{cosbkrn} and the fate of black hole collapse \cite{bhbkrn},
and the study of symmetry breaking and critical phenomena via interacting
field theory in curved spacetime (see, e.g., \cite{HuOC}).
These studies are needed for understanding the dynamics of
the inflationary universe, such as spinodal decomposition and nucleation,
and from it, astrophysical consequences such as entropy generation and
galaxy formation. Since particle creation stems from amplification of
vacuum fluctuations, and galaxy formation have them as seeds, both directions
share a common need to understand better the properties of vacuum fluctuations
of quantum fields and how they affect the dynamics of spacetime in the
early universe and state of the matter at the observable
classical domains or epochs. In the 90's, the theoretical underpinning of this
issue was taken up afresh in the work of Ford and coworkers \cite{Ford,KuoFord},
who investigated amongst other problems the effect of vacuum fluctuations
in a quantum
field on the causal structure of a quantum field theory, and in the work of
Hu and coworkers \cite{HuTsukuba,CH94}, who introduced
non-equilibrium statistical mechanical concepts and techniques for the study
of noise and fluctuations associated with a quantum field, with applications
to backreaction problems in semiclassical gravity \cite{HM3,fdrsc} and
structure formation problems in inflationary cosmology \cite{HuBelgium,CH95}.

In this paper, we aim to set up the basic framework for the calculation of
the fluctuations of energy momentum for quantum fields in curved spacetimes,
focussing on the generalization of the zeta-function regularization method
\cite{EliEtAl}.
Elegant and powerful, its application is unfortunately limited to spacetimes
which admit an Euclidean section. But this includes many cases of physical
importance, such as the de Sitter universe (in the $S^4$ coordinatization),
Kaluza Klein theory ($M^4 \times B^D$), and finite temperature theory
(in the imaginary-time formulation $R^3 \times S^1$). Later we intend to
connect it with the point-separation method, which, as one of us has 
anticipated,
actually contains untapped useful information about the statistical mechanical
(or kinetic theory) properties of systems of particles interacting with fields 
in a curved spacetime.
The result obtained here is also useful for the calculation
of fluctuations of Casimir energy relevant in atom-optics \cite{Barton},
mesoscopic physics \cite{meso}, and correlations of quantum fields \cite{cddn,Erice}
relevant to black hole fluctuations \cite{BekMuk,HPR,HRS} and possible Planck scale
phase transitions \cite{HuBanff}.

The goal of this paper is to develop a regularized expression for the stress
energy tensor two-point function of a quantum field in a curved spacetime.
This is done by first expressing the two-point function solely in terms of
the quantum field effective action and its variations with respect to the
metric. The effective action is derived by using the trace of the heat kernel
that corresponds to the field action. Since the effective action can be
expressed as a function of the Mellin transform of the heat kernel, the
needed variations of the effective action can  be evaluated by
considering the variations of the heat kernel.
The variations of the heat kernel can be related to variations of
an operator acting on the heat kernel in the varied heat equation.

In Section II we relate the stress energy two point function to the
second metric variation of the effective action. In Section III
we show how to express this second variation in terms of the generalized
zeta function for the system. Section IV presents the specialization of
these expressions to the Klein-Gordon scalar field. In particular, we
derive the expression for the variance of the energy density, and discuss
the need for regularization. In Secs. V and VI we explicitly compute
the energy density variance for a flat spacetime with one periodic dimension 
and for the Einstein universe respectively. We give a brief discussion
of our results in Sec. VII.

\section{Stress energy two-point function}

        We start by considering the generating functional (or partition
function) of a scalar field $\phi$ in the Euclidean section $\Sigma$
of a spacetime manifold $\cal M$,
\begin{equation}
Z=\int {\cal D}\phi\; e^{-S[\phi]} = \langle 0,{\rm out}|0,{\rm in}\rangle 
\end{equation}
and its functional derivatives with respect to the metric:
\begin{eqnarray}
\frac{\delta Z}{\delta g^{ab}(x)} &=& -\int{\cal D}\phi \frac{\delta S}
                                                     {\delta g^{ab}(x)}e^{-S} 
            = -\langle 0,{\rm out}|\frac{\delta S}{\delta g^{ab}(x)}|0,{\rm in}
                                                                  \rangle ,\cr
\frac{\delta^2 Z}{\delta g^{ab}(x)\delta g^{cd}(y)} &=&
     \int{\cal D}\phi\left[ \frac{\delta S}{\delta g^{ab}(x)}
                                           \frac{\delta S}{\delta g^{cd}(y)} - 
     \frac{\delta^2 S}{\delta g^{ab}(x)\delta g^{cd}(y)}\right] e^{-S}\cr
&=&\langle 0,{\rm out}|\frac{\delta S}{\delta g^{ab}(x)}
                            \frac{\delta S}{\delta g^{cd}(y)}|0,{\rm in}\rangle  
          - \langle 0,{\rm out}|
          \frac{\delta^2 S}{\delta g^{ab}(x)\delta g^{cd}(y)}|0,{\rm in}\rangle 
\end{eqnarray}
where $|0 in, out>$ are the vacuua defined at the in and out states.
In terms of the effective action $W = \log Z$, this becomes
\begin{equation}
\frac{\delta^2 W}{\delta g^{ab}(x)\delta g^{cd}(y)} 
=  \frac{1}{Z}\frac{\delta^2 Z}{\delta g^{ab}(x)\delta g^{cd}(y)} 
- \frac{\delta W}{\delta g^{ab}(x)}\frac{\delta W}{\delta g^{cd}(y)}.
\end{equation}

The expectation value of the quantum  stress energy tensor is given by
\begin{equation}
\langle T_{ab}\rangle = \frac{\langle 0,{\rm out}|T_{ab}|0,{\rm in}\rangle }
                                       {\langle 0,{\rm out}|0,{\rm in}\rangle }
=-\frac{2}{\sqrt{g(x)}}\frac{\langle 0,{\rm out}|\delta S/\delta g^{ab}(x)
                   |0,{\rm in}\rangle }{\langle 0,{\rm out}|0,{\rm in}\rangle }
= -\frac{2}{\sqrt{g(x)}}\frac{\delta W}{\delta g^{ab}(x)}.
\label{eq:quan-stress-energy}
\end{equation}
In analogue with this, we define the 
correlation function for the stress energy tensor as
\begin{eqnarray}
\langle T_{ab}(x)T_{cd}(y)\rangle &=& 
           \frac{\langle 0,{\rm out}|T_{ab}(x)T_{cd}(y)|0,{\rm in}\rangle }
                                   {\langle 0,{\rm out}|0,{\rm in}\rangle }\cr
    &=& \frac{4}{\sqrt{g(x)g(y)}}\frac{1}{Z}
          \langle 0,{\rm out}|\frac{\delta S}{\delta g^{ab}(x)}
                       \frac{\delta S}{\delta g^{cd}(y)}|0,{\rm in}\rangle \cr
    &=& \frac{4}{\sqrt{g(x)g(y)}}\left[  
      \frac{\delta^2 W}{\delta g^{ab}(x)\delta g^{cd}(y)} + 
             \frac{\delta W}{\delta g^{ab}(x)}\frac{\delta W}{\delta g^{cd}(y)}
           + \langle \frac{\delta^2 S}{\delta g^{ab}(x)\delta g^{cd}(y)} \rangle
    \right].
\end{eqnarray}
For any local action the last term will not contribute to the final expression
for the stress-energy correlation function. For such an action,
 this expression will depend on $x,y$ as a delta function $\delta(x-y)$.
Thus it need not be considered when computing the correlation function for
$x\ne y$. The autocorrelation is understood as resulting from this by taking
the $y\rightarrow x$ or coincidence limit.
Recognizing the second term in the last line as a product of expectation values
of the stress-energy tensor, we can define the bitensor

\begin{equation}
\Delta T^2_{abcd}(x,y) \equiv
\langle T_{ab}(x)T_{cd}(y)\rangle - 
                \langle T_{ab}(x)\rangle\langle T_{cd}(y)\rangle =
\frac{4}{\sqrt{g(x)g(y)}}
\frac{\delta^2 W}{\delta g^{ab}(x)\delta g^{cd}(y)}.
\end{equation}

\section{Second variation of the effective action}

The classical action of a scalar field $\phi (x)$ is
\begin{equation}
S[\phi] = \frac{1}{2} \int d{\bf x} \phi(x) H \phi(x),
\end{equation}
where $H$ is a second order elliptic operator.
From the spectral decomposition of this operator,
$H=\sum_n \lambda_n |n\rangle\langle n|$, where $n$ denotes 
the collective indices of the spectrum, the effective
action can be expressed as
\begin{equation}
W = -\frac{1}{2} \ln \det (H/\mu) 
  = -\frac{1}{2} {\rm Tr}\; \ln \frac{H}{\mu} = 
                                  -\frac{1}{2}\sum_n \ln\frac{\lambda_n}{\mu},
\end{equation}
where we assume the zero modes of $H$ have been projected
out and $\mu$ is a normalization  constant with dimensions of mass squared.
This expression for the effective action is only
formal since $H$ is not trace class. We regularize
the effective action and the expressions derived from
it via the zeta function method \cite{DowCri,Haw}.

The zeta function for this system is defined as
\begin{equation}
\zeta_H(s) = {\rm Tr}\;  e^{-s\ln H/\mu} = \mu^s \sum_n \lambda_n^{-s}.
\end{equation}
For $2s>{\rm dim}\;{\cal M}$, this sum is convergent. Then
an analytic continuation in $s$ is found such that it includes a neighborhood of
$s=0$. From this analytic continuation the regularized 
effective action is given as
\begin{equation}
W_R = \left. \frac{1}{2}\frac{d\zeta(s)}{ds}\right|_{s=0}
= \frac{1}{2}\zeta'(0).
\end{equation}
For positive $H$ the definition of the gamma function yields
\begin{equation}
\zeta_H(s) = \frac{\mu^s}{\Gamma\left(s \right)}\int_0^\infty t^{s-1}
{\rm Tr}\;  U_0(t) dt,\quad U_0(t) = e^{-t H},
\label{eq:zeta1}\end{equation}
i.e. the zeta function is given as the Mellin transform of the
trace of the heat kernel $U_0(t)$.
We know ${\rm Tr}\;  U_0(t) \sim t^{-d/2}$ for $t\rightarrow 0$.
Hence for \ (\ref{eq:zeta1}) to be convergent, we have the
condition $s > d/2$. This is most often the expression
from which an analytic continuation is derived. In fact,
 one of the main points of the zeta function
idea is that formal expressions such as ${\rm Tr}\;  U_0(t)$ need 
be modified by the introduction of a factor $t^\nu$. Then
once the analytic continuation is found, one takes $\nu = 0$.

We now consider the effect of two small metric perturbations
$\delta_1$ and $\delta_2$ on the effective
action.(They are assumed to be
independent of the order with which they act).
The response of the effective action to these perturbations is
\begin{eqnarray}
\delta_2\delta_1 W_R[g]
&=&W_R[g+\delta_1 + \delta_2]+W_R[g]-W_R[g+\delta_1]-W_R[g+\delta_2]\cr
&=&\left. \frac{1}{2}\frac{d}{ds} \left( 
      \delta_2 \delta_1 \zeta_H(s) \right) \right|_{s=0},
\label{eq:2nd-var-W}\end{eqnarray}
\begin{equation}
\delta_2 \delta_1 \zeta_H(s) = \frac{\mu^s}{\Gamma\left(s \right)}
\int_0^\infty \!\!dt\;t^{s-1}{\rm Tr}\; \left\{
   e^{-t\left(H + \delta_1 H + \delta_2 H\right)} 
 - e^{-t\left(H + \delta_1 H\right)}- e^{-t\left(H + \delta_2 H\right)}
 + e^{-t H}\right\}
\end{equation}
To evaluate this, we use the Schwinger perturbative expansion
\cite{Schwinger}. For $U(t) = e^{-t\left(H+H_1\right)}$, where $H_1 = \delta_1 H
\ll H$,
\begin{equation}
{\rm Tr}\;  U(t) = {\rm Tr}\;  U_0(t) - t {\rm Tr}\; \left[H_1 U_0(t)\right]
+\frac{t^2}{2}\int_0^1 du_1
{\rm Tr}\; \left[H_1 U_0((1-u_1)t) H_1 U_0(u_1 t)\right]+ \cdots
\end{equation}
Using this, we can write the response of the zeta function
to these perturbations as
\[
\delta_2\delta_1 \zeta_H(s) =\frac{\mu^s}{2\Gamma\left(s \right)}
\int_0^\infty\!\!dt\;t^{s+1}\int_0^1\!\! du_1
\left\{
{\rm Tr}\; \left[(\delta_1 H)U_0((1-u_1)t)(\delta_2 H)U_0(u_1 t)\right]\right.
\]
\begin{equation} \left.
+{\rm Tr}\; \left[(\delta_2 H)U_0((1-u_1)t)(\delta_1 H)U_0(u_1 t)\right]
\right\}.
\end{equation}
As it stands, it is not finite. When the traces
are taken involving $U_0((1-u_1)t)$ and $U_0(u_1 t)$, the
divergences at $(1-u_1)t,u_1 t \rightarrow 0$ are present.
At this point, we modify the above expression for the
second variation of the zeta function by introducing
the factor $\left[u_1(1-u_1)t^2\right]^\nu$. At the
end of the calculation, once the analytic continuation is
found, we set $\nu=0$. We view the introduction of this
factor as an extention of the zeta function method to
the situation where the second variation is needed. The replacement of
$U_0(t) \rightarrow t^\nu U_0(t)$ is the spirit of the usual
zeta function method and reproduces the usual
results when applied to the traditional problems, such as finding the
first variation, which produces the expectation value of the  quantum
stress energy tensor.
After the change of variables
\begin{equation}
u = (1-u_1)t, \quad v = u_1 t
\end{equation}
the twice varied zeta function transforms to
\begin{equation}
\delta_2\delta_1 \zeta_H(s) =
\frac{\mu^s}{2\Gamma\left(s \right)}\int_0^\infty\!\!du\int_0^\infty\!\!dv
(u+v)^s (uv)^\nu \left\{
{\rm Tr}\; \left[(\delta_1 H)U_0(u)(\delta_2 H)U_0(v)\right]
+{\rm Tr}\; \left[(\delta_2 H)U_0(u)(\delta_1 H)U_0(v)\right]
\right\}.
\end{equation}

Considering the first trace in the above expression, we find
\begin{eqnarray}
{\rm Tr}\; \left[(\delta_1 H)U_0(u)(\delta_2 H)U_0(v)\right]
&=& \sum_{n,n'} \langle n'|(\delta_1 H)e^{-u H}|n\rangle\langle n|
(\delta_2 H)e^{-v H}|n'\rangle\cr
&=& \sum_{n,n'} e^{-u \lambda_n -v \lambda_m'}
\langle n'|(\delta_1 H)|n\rangle\langle n|(\delta_2 H)|n'\rangle.
\end{eqnarray}
By defining
\begin{equation}
T_{ab}[\phi_n(x),\phi_n'^*(x)] \equiv 
-\frac{2}{\sqrt{g(x)}}\langle n'|\frac{\delta H}{\delta g_{ab}(x)}|n\rangle
= -\frac{2}{\sqrt{g(x)}}\int d{\bf z}
\phi_n'^*(z)\frac{\delta H}{\delta g_{ab}(x)}\phi_n(z),
\label{eq:my-se-tensor}
\end{equation}
we can now write the stress-energy correlation bitensor
as
\begin{equation}
\Delta T^2_{abcd}(x,y) = \frac{1}{2}\frac{d}{ds}\left[
\frac{\mu^s}{\Gamma\left(s \right)}\int_0^\infty\!\!du\int_0^\infty\!\!dv
(u+v)^s (uv)^\nu \sum_{n,n'} e^{-u \lambda_n - v \lambda_{n'}}
T_{ab}[\phi_n(x),\phi_{n'}^*(x)]T_{cd}[\phi_{n'}(y),\phi_n^*(y)]
\right]
\end{equation}
where the $s,\nu\rightarrow 0$ limit is understood.

For the rest of this paper, we will specialize to the cases of homogeneous
spacetimes. This implies $\Delta T^2_{abcd}(x,y)$ will only depend
on $r=x-y$. We find it convenient to average over all $x$. We can do this
while leaving the points separated by $r$. Also, the homogeneity will usually
lead to a degeneracy of the eigenvalues. Thus the collective quantum number
$n$ can be split into principal and degenerate parts $n \rightarrow n,m$
and the eigenvalues only depend on $n$:
$\lambda_{nm}\rightarrow\lambda_{n}$. This allows the sum over the degenerate
indices to be done before evaluation of the zeta function.
Putting all these together the stress energy two point function becomes
\[
\Delta T^2_{abcd}(r) = \frac{1}{2\Omega}\frac{d}{ds}\left[
\frac{\mu^s}{\Gamma\left(s \right)}\int_0^\infty\!\!du\int_0^\infty\!\!dv
(u+v)^s (uv)^\nu \right.
\]
\begin{equation}
\times\left.
\left( \sum_{n,n'} e^{-u \lambda_n - v \lambda_{n'}}
\sum_{mm'}\int_{\cal M}\!\!d{\bf x}\;
T_{ab}\left[\phi_{nm}(x),\phi_{n'm'}^*(x)\right]
T_{cd}\left[\phi_{n'm'}(x+r),\phi_{nm}^*(x+r)\right]
\right)\right],
\end{equation}
where $\Omega = \int_{\cal M}d{\bf x}$, the volume of the manifold. If the
manifold is noncompact, it is understood to be the unit volume.

\section{Form for the Klein-Gordon field}

We now develop the general form for the second variation of the zeta function
for the Klein-Gordon field. In the Lorentzian sector, we use the MTW signature
convention $(-1,1,\ldots)$, and thus in the Euclidean sector, we have the
signature $(1,1,\ldots)$. We assume the metric can be given
the form
\begin{equation}
g_{ab} = \left( \begin{array}{cc}
1 & 0\cr 0 & h_{ij}
\end{array} \right),
\end{equation}
where $h_{ij}$ is the metric for the spatial section. We denote the time and
spatial variables by $x=(\tau,{\bf x})$, the invariant spatial volume
form by $d{\bf x}$, and the spatial manifold by $\Sigma$ (thus
${\cal M} = S^1\times\Sigma$).

The wave operator for the Klein-Gordon field is given by
\begin{equation}
H = -\Box + \xi R + m^2 = -\frac{\partial^2}{\partial\tau^2}
- {}^{\Sigma}\!\Delta   + \xi R + m^2.
\end{equation}
Let $u_n({\bf x})$ be the eigenfunctions of ${}^{\Sigma}\!\Delta $: 
${}^{\Sigma}\!\Delta  u_n({\bf x}) = -\kappa^2_n u_n({\bf x})$,
where $n$ denotes the (collective) quantum numbers for the spatial part of the
spectrum.
We assume the $u_n({\bf x})$ are orthonormal. The Euclidean time is made
periodic with a peroid of $\beta=1/T$, where $T$ can be interpreted as a
temperature. The eigenfunctions are thus given by
\begin{mathletters}
\label{eq:KG-modes}
\begin{equation}
\phi_{k_0,n}(x) = \frac{e^{-ik_0 \tau}}{\sqrt{\beta}}u_n({\bf x}),
\quad k_0 = \frac{2\pi n_0}{\beta},\quad n_0  = 0,\pm 1,\pm 2,\ldots
\end{equation}
and the eigenvalues by  (see e.g., \cite{OHS})
\begin{equation}
\lambda_{k_0,n} = k_0^2 + \kappa_n^2 + \xi R + m^2.
\end{equation}
\end{mathletters}

From our definition of the stress-energy tensor\ (\ref{eq:my-se-tensor})
we find
\begin{eqnarray}
T_{ab}[\psi,\phi](x) &\equiv& \frac{2}{\sqrt{g(x)}}\int\!\!\sqrt{g(x')}
\psi(x')\left(\frac{\delta H_{x'}\phi(x') }{\delta g^{ab}(x)} \right) dx'\cr
&=& - 2\nabla_a \psi \nabla_b \phi + g_{ab}\left(
\nabla_c \psi \nabla^c \phi + \psi\nabla_c  \nabla^c \phi \right)
+ 2\xi \psi\phi R_{ab}.\label{eq:Tmunu}
\end{eqnarray}
This differs from the usual definition of $T_{ab}[\psi,\phi]$, but
this is to be expected. The eigenvalues used are themselves different,
since we have the extra $\tau$-degree of freedom.

In this paper, we wish to concentrate on computing the autocorrelation of the
energy density 
\begin{equation}
\Delta\rho^2(x)\equiv \lim_{y\rightarrow x}\left(
\langle\rho(x)\rho(y)\rangle-\langle\rho(x)\rangle \langle\rho(y)\rangle\right),
\quad \rho = T_{00}.
\end{equation}
This will give a measure for the magnitude of the fluctuations of the
stress-energy vacuum expectation value. This choice yields
\begin{equation}
\rho[\phi,\psi] = -\left(\partial_\tau\phi \partial_\tau\psi
-\psi \partial^2_\tau \phi\right)
+\left( \nabla_i \psi \nabla^i \phi + \psi {}^{\Sigma}\!\Delta  \phi\right),
\end{equation}
where $i$ is summed over the spatial degrees of freedom only.
This becomes
\begin{equation}
\rho[\phi_n,\phi^*_{n'}] = - \left(k_0^2 + k_0 k'_0 + \kappa_n^2\right)
\phi_n\phi^*_{n'} 
+ \left(\nabla_i \phi_n\right)\left( \nabla^i \phi^*_{n'}\right).
\end{equation}
when we specialize to eigenmodes \ (\ref{eq:KG-modes}).
When we consider $|\rho[\phi_n,\phi^*_{n'}]|^2$ the contribution from the
$\phi^*_n\phi_{n'} \nabla_i \phi_n \nabla^i \phi^*_{n'}$ and its conjugate
will vanish when summed over. This will be shown case by case. 
This assumption provides
\begin{equation}
\left|\rho[\phi_n,\phi^*_{n'}]\right|^2 
= \left(k_0^2 + k_0 k'_0 + \kappa_n^2\right)
\left({k'_0}^2 + k_0 k'_0 + \kappa_{n'}^2\right)|\phi_n|^2|\phi^*_{n'}|^2
 + \left|
\left(\nabla_i \phi_n\right)\left( \nabla^i \phi^*_{n'}\right)\right|^2.
\end{equation}

At this point, we wish to take the zero temperature limit, i.e. $\beta
\rightarrow \infty$ whereby $\sum_{n_0=-\infty}^\infty \rightarrow
(\beta/2\pi)\int_{-\infty}^\infty dk_0$. We can now do these sums/integrals
easily, since they amount to calculating the moments of gaussians. We find
\begin{eqnarray}
\sum_{n_0=-\infty}^\infty e^{-(2\pi n_0/\beta)^2u} &\rightarrow&
\frac{\beta}{2\pi}\int_{-\infty}^\infty e^{-uk_0^2} dk_0 = 
\frac{\beta}{2\sqrt{\pi}}u^{-1/2}\cr
\sum_{n_0=-\infty}^\infty \left(\frac{2\pi n_0}{\beta}\right)^2
e^{-(2\pi n_0/\beta)^2u} &\rightarrow&
\frac{\beta}{2\pi}\int_{-\infty}^\infty k_0^2 e^{-uk_0^2} dk_0 = 
\frac{\beta}{4\sqrt{\pi}}u^{-3/2}
\end{eqnarray}
and there is zero contribution for any odd power of $k_0$ or $k'_0$. The
normalization of $\phi_n$ and $\phi_{n'}$ provide a factor of $\beta^{-2}$ which cancels
the powers of $\beta$ introduced in approaching the $\beta\rightarrow\infty$
limit. Introducing the functions
\begin{mathletters}
\label{eq:define-XTP}
\begin{eqnarray}
\Xi_{nn'}(u,v) &=& \frac{1}{8\pi\Omega (uv)^\frac{1}{2}}\left\{
\frac{1}{uv} + \frac{\kappa_n^2}{v} + \frac{\kappa_{n'}^2}{u}
+ 2\kappa_n^2\kappa_{n'}^2\right\}
\sum_{mm'}\int_\Sigma  |u_{nm}|^2 |u_{n'm'}|^2 d{\bf x} \label{eq:define-Xi}\\
\Theta_{nn'}(u,v) &=& \frac{1}{8\pi\Omega (uv)^\frac{1}{2}}\left(
\frac{1}{u}+2\kappa_n^2\right) \sum_{mm'}
\int_\Sigma u_{nm}\nabla_i u^*_{nm} u^*_{n'm'}\nabla^i u_{n'm'}d{\bf x}
\label{eq:define-Theta}\\
\Pi_{nn'}(u,v) &=& \frac{1}{4\pi\Omega (uv)^\frac{1}{2}}
\sum_{mm'}\int_\Sigma 
\left|\nabla_i u_{nm}\nabla^i u^*_{n'm'}  \right|^2 d{\bf x}\label{eq:define-Pi}
\end{eqnarray}
\end{mathletters}
and
\begin{equation}
\Psi_{nn'}(u,v) = \Xi_{nn'}(u,v) + \Theta_{nn'}(u,v)
 + \Theta_{n'n}(v,u) + \Pi_{nn'}(u,v), \label{eq:define-Psi}
\end{equation}
we define the zeta function 
\begin{equation}
\zeta_\Psi(s,\nu) = 
\sum_{n,n'}
\int_0^\infty\!\!\int_0^\infty\!\! (u+v)^s (uv)^\nu \Psi_{nn'}(u,v) 
e^{-\lambda_n u -\lambda_{n'}v}dvdu. \label{eq-zeta_tilde}
\end{equation}
From it, we derive an expression for the regularized
autocorrelation of the energy density as
\begin{equation}
\Delta\rho^2 = \frac{1}{2}\left.
\frac{d}{ds}\left[\frac{\mu^s}{\Gamma\left(s \right)}\zeta_\Psi(s,\nu)
                                                     \right]\right|_{s=0,\nu=0}
= \frac{1}{2} \lim_{\nu\rightarrow 0} \zeta_\Psi(0,\nu) \label{eq:Delta-rho}
\end{equation}
where we have used $1/\Gamma\left(s \right)\sim s+\gamma s^2 +0(s^3)$
($\gamma$ is Euler's constant) for $s\sim 0$. 
We have assumed $\nu > d/2 + 1$, in which case both $\zeta_\Psi(0,\nu)$
and $d\zeta_\Psi(s,\nu)/ds|_{s=0}$ are finite. Thus to find the regularized
expression for $\Delta\rho^2$, we need to find the analytic continuation
of $\zeta_\Psi(0,\nu)$ in $\nu$ such that it is finite at $\nu=0$.

\section{Fluctuations for $\Sigma = R^d \times S^1$}

As the first application, we calculate in this section the variance of the
energy density for a massless ($m=0$) minimally coupled ($\xi=0$) scalar field
on a $(d+2)$-dimensional flat ($R=0$) spacetime which is periodic on one
spatial dimension with period $L$. For this geometry, the spatial eigenfunctions are
\begin{equation}
u_{{\bf k} n}({\bf x},z) = \frac{e^{i{\bf k}\cdot{\bf x} + i l z}}
                                                            {\sqrt{(2\pi)^d L}},
\quad {\bf k} \in {\bf R}^d,\quad n=0,\pm 1, \pm 2,\ldots,
\end{equation}
with $l = 2\pi/L$.
Denoting by ${\bf x}=(x_1,\ldots,x_d)$ the coordinates for the open dimensions
and $z$ for the one compact dimension, we have
${}^{\Sigma}\!\Delta  = \sum_{j=1}^d \partial^2_{x_j} + \partial^2_z$ and hence
$\kappa^2_{k,n} = k^2 + l^2 n^2$, ($k^2 = |{\bf k}|^2$). From this we see we
should take $k$ and $n$ as the principal indices and the angular degeneracy
of ${\bf k}$ as the degenerate indices, i.e. $\sum_m \rightarrow \int d\Omega_{d-1}$,
integration over the unit $d\!-\!1$ sphere. Where there is no confusion, we will
denote the volume of the sphere as $S^{d-1}$.

To evaluate \ (\ref{eq:define-XTP}) we sum over the 
degenerate indices and perform the volume average
\begin{equation}
\int d\Omega_{d-1}\int d\Omega'_{d-1}\frac{1}{{\rm Vol}(\Sigma)}\int d{\bf x} dz
|u_{k n}|^2 |u_{k' n'}|^2 = \left(\frac{S^{d-1}}{(2\pi)^d L}\right)^2
\end{equation}
from which \ (\ref{eq:define-Xi}) becomes
\begin{equation}
\Xi_{knk'n'}(u,v) = \frac{1}{8\pi}\left(\frac{S^{d-1}}{(2\pi)^d L}\right)^2
\frac{1}{(uv)^\frac{1}{2}}\left\{
\frac{1}{uv} + \frac{k^2 + l^2 n^2}{2v} +
\frac{k'^2 + l^2 n'^2}{2u} +
2\left(k^2 + l^2n^2 \right)
\left(k'^2 + l^2n'^2 \right)
\right\}
\end{equation}
Also, since $\nabla_i u_{kn} \nabla^i u^*_{k'n'} =
(kk'\cos\gamma + l^2 n n')u_{kn}u^*_{k'n'}$ where
$\cos\gamma = {\hat {\bf k}}\cdot{\hat {\bf k}'}$, the
momentum correlation term \ (\ref{eq:define-Pi}) is
\begin{equation}
\Pi_{knk'n'}(u,v) = \frac{1}{8\pi}\left(\frac{S^{d-1}}{(2\pi)^d L}\right)^2
\frac{2}{(uv)^\frac{1}{2}}\left\{
\frac{k^2 k'^2}{d} + l^4 n^2 n'^2 \right\}.
\end{equation}
Here we have used $\int d\Omega_{d-1}\int d\Omega'_{d-1}\cos\gamma= 0$
and \ (\ref{eq:beta}) from the appendix:
$\int d\Omega_{d-1}\int d\Omega'_{d-1}\cos^2\gamma=(S^{d-1})^2/d$.
Also, when summed over $n$ and $n'$, $\Theta_{knk'n'}$
vanishes.

We now turn to the principal index sums for $k$ and $k'$.
First consider the case when $n=0$ (or $n'=0$), this leads us to
evaluate
\begin{equation}
\int_0^\infty\!\!dk\;k^{d-1}\int_0^\infty\!\!du\;u^{\nu-a} k^b e^{-k^2 u}.
\label{eq:n=0-term}\end{equation}
Here $(a,b) = (3/2,0)$ or $(1/2,2)$, and $2a+b-3=0$. We assume
a boundary on the non-periodic dimensions at large distance and
moved to infinity. This is effected by having
$\int_0^\infty dk \rightarrow \lim_{\epsilon\rightarrow 0}\int_\epsilon^\infty$.
Then \ (\ref{eq:n=0-term}) becomes
\begin{eqnarray}
\int_\epsilon^\infty\!\!dk\;k^{d-1}\int_0^\infty\!\!du\;u^{\nu-a} 
k^b e^{-k^2 u} &=&
\Gamma\left(\nu-a+1 \right)\int_\epsilon^\infty\!\!dk\;k^{d-2\nu}\cr\cr
&\stackrel{\nu\rightarrow 0}{\longrightarrow}&  
-\frac{\Gamma\left(1-a \right)}{d+1}\epsilon^{d+1}
\stackrel{\epsilon\rightarrow 0}{\longrightarrow} 0
\end{eqnarray}
where we have used $\nu>d/2$ in evaluating the $k-$integration. Thus we find
there is no contribution from the $n=0$ or $n'=0$ terms in $\zeta_\Psi$.

Turning to the case where $n$ and $n'$ do not vanish,
we consider the function
\begin{equation}
\Psi_{nn'}(u,v) = \int_0^\infty k^{d-1}dk \int_0^\infty k'^{d-1}dk'\;
\Psi_{knk'n'}(u,v)\; e^{-u k^2 - v k'^2}.
\end{equation}
In the integrand, there are terms with either factors of $k^{d-1}$ or $k^{d+1}$,
similiarly for $k'$. Using
\begin{equation}
\int_0^\infty k^{d-1}dk e^{-u k^2} = \frac{1}{2}\Gamma\left(\frac{d}{2} 
                                                               \right)u^{-d/2}
\quad{\rm and}\quad
\int_0^\infty k^{d+1}dk e^{-u k^2} = \frac{1}{2}\Gamma\left(\frac{d}{2} 
                                                               \right)u^{-d/2}
\frac{d}{2u},
\end{equation}
we have the rule that upon doing the $k,k'$ integrations, we get an
overall factor of $(\Gamma\left(\frac{d}{2} \right)/2)^2(uv)^{-d/2}$ and each
factor of $k^2$ becomes $d/2u$, and $k'^2$ becomes $d/2v$. Applying this rule,
\ (\ref{eq:define-Psi}) has been determined
\begin{equation}
\Psi_{nn'}\left(\frac{u}{l^2},\frac{v}{l^2}\right) =
\frac{1}{8\pi}\left[
\frac{S^{d-1}l^{d+3}}{2(2\pi)^d L}\Gamma\left(\frac{d}{2} \right)\right]^2
\left(\frac{1}{uv}\right)^{\frac{d+3}{2}}\left\{
\frac{(d+1)(d+2)}{2} + (1+d)(un^2 + vn'^2) 
+4uv n^2 n'^2\right\}
\end{equation}
and the coresponding zeta function is 
\begin{equation}
\zeta_\Psi(0,\nu) = B\sum_{n,n'=1}^\infty
\int_0^\infty\!\!du \int_0^\infty\!\!dv (uv)^{\nu -\frac{d+3}{2}}
\left\{ \frac{(d+1)(d+2)}{2} + (1+d)(un^2 + vn'^2) 
+4uv n^2 n'^2\right\} e^{-n^2u-n'^2v},
\end{equation}
where
\begin{equation}
B = \frac{1}{8\pi}\left[
\frac{S^{d-1}l^{d+1-2\nu}}{(2\pi)^d L}\Gamma\left(\frac{d}{2} \right)\right]^2
\end{equation}

This form of the zeta function now allows us to perform the needed analytic
continuation. We make use of the relation 
\begin{equation}
\sum_{n=1}^\infty n^a \int_0^\infty\! t^{s-1} e^{-n^2t}
=\Gamma\left(s \right)\zeta_R(2s-a),\label{eq:Riemann-zeta}
\end{equation}
where $\zeta_R(s)$ is the Riemann zeta function. Recalling 
\ (\ref{eq:Delta-rho}), the variance of the energy density is
\begin{equation}
\Delta\rho^2 = \frac{B}{2}
\zeta_R(-(d+1))^2\left\{
\frac{(d+1)(d+2)}{2}\Gamma\left(-\frac{d+1}{2} \right)^2
+2(d+1)\Gamma\left(-\frac{d+1}{2} \right)\Gamma\left(-\frac{d-1}{2} \right)
+4\Gamma\left(-\frac{d-1}{2} \right)^2
\right\}.
\end{equation}
Since $\Gamma\left(-(d-1)/2 \right) = -((d+1)/2)\Gamma\left(-(d+1)/2 \right)$,
the second and third terms in the above expression cancel,
leaving
\begin{equation}
\Delta\rho^2 =  \frac{(d+1)(d+2)}{2}\left[
\frac{S^{d-1}}{2\pi^{d+1}L^{d+2}}
\Gamma\left(\frac{d}{2}\right)\Gamma\left(\frac{d+2}{2}\right)
\zeta_R(d+2) \right]^2
\end{equation}
by way of the Riemann zeta function reflection formula:
\begin{equation}
\Gamma\left(\frac{s}{2} \right)\zeta_R(s) = \pi^{s-\frac{1}{2}}
                                 \Gamma\left(\frac{1-s}{s} \right)\zeta_R(1-s).
\label{eq-Rzeta-reflect}
\end{equation}
 This is our result for the
zero temperature variance of the vacuum energy density for a 
massless minimally-coupled quantum scalar field on a $d+2$-dimensional spacetime 
periodic in one spatial dimension. To get a measure of the 
fluctuations, we consider the dimensionless quantity
\begin{equation}
\Delta' = \frac{\langle\rho^2\rangle-\langle\rho\rangle^2}
{\langle\rho\rangle^2}. \label{eq:define-Delta'}
\end{equation}
For the system at hand, the energy density is \cite{EliEtAl}:
\begin{equation}
\langle\rho\rangle = -\frac{S^{d-1}}{2\pi^{d+1}L^{d+2}}
\Gamma\left(\frac{d}{2}\right)\Gamma\left(\frac{d+2}{2}\right)
\zeta_R(d+2).
\end{equation}
Thus we get one of the main results of this paper:
\begin{equation}
\Delta'(\Sigma=R^d\times S^1) = \frac{(d+1)(d+2)}{2}.
\end{equation}

Kuo and Ford \cite{KuoFord} computed the
same measure of the fluctuations for the case of 
$\Sigma=R^2\times S^1$ via a different method.
Their result of $\Delta'=6$ is obtained here for $d=2$.
It is interesting to note
that the relative amount of fluctuation increases quadratically
with the dimension of the spacetime.

\section{Fluctuations for $\Sigma=S^3$}

As a second example, we calculate the fluctuations of energy
density for a massless ($m=0$) conformally coupled ($\xi=1/6$)
scalar field on a 3-dimensional space of constant curvature $S^3$ with radius
$a$. The spacetime $\cal M$ is then the Einstein Universe.

We start by writing the spatial metric as \cite{Hu73}
\begin{equation}
ds^2 = \gamma_{ab}\sigma^a\sigma^b = \sum_{a=1}^3 l_a^2(\sigma^a)^2,
\end{equation}
where the $\sigma^a$'s are the basis one-forms on the three sphere satisfying
the structure relations
\begin{equation}
d\sigma^a = \frac{1}{2} \epsilon^a_{bc}\sigma^b\sigma^c.
\end{equation}
Here $\epsilon^a_{bc}$, components of the totally antisymmetric tensor,
are the structure constants for the rotation group
$SO(3)$. The $\gamma_{ab}=l_a^2\delta_{ab}$ are constants of the space
(the principal curvature radii in a static mixmaster universe \cite{Misner69})
and for the Einstein Universe $\Sigma = S^3$ we have $l_1=l_2=l_3=a/2$. The
curvature scalar is $R=6/a$ and the volume is $\Omega=2\pi^2 a^3$.

Using the Euler angle parametrization, the basis forms are 
\begin{equation}
\begin{array}{rcrrcr}
\sigma^1 &=&         &-\sin\psi d\theta &+& \cos\psi\sin\theta d\phi,\\
\sigma^2 &=&         & \cos\psi d\theta &+& \sin\psi\sin\theta d\phi,\\
\sigma^3 &=&   d\psi &         &+& \cos\theta d\phi,
\end{array}
\end{equation}
$$0\le\theta\le\pi,\quad 0\le\phi\le 2\pi,\quad 0\le\psi\le 4\pi.$$
Taking ${\bf e}_a$ as the invariant vectors dual to $\sigma^a$ and
defining the angular momentum operators $L_a = i{\bf e}_a$, the spatial
Laplacian becomes
\begin{equation}
{}^{\Sigma}\!\Delta  = \sum_{a=1}^3 l_a^{-2}({\bf e}_a)^2
=-\frac{4}{a^2}\left(L_1^2 + L_2^2 + L_3^2\right)=-\frac{4L^2}{a^2}.
\end{equation}
For the harmonic functions on $S^3$ we can use the SO(3) representation (Wigner)
functions $D^J_{KM}(\theta,\psi,\phi)$, where
$J=0,\frac{1}{2},1,\frac{3}{2},\ldots$ are the principal quantum numbers and
$K,M = -J,-J+1,\ldots,J-1,J$ are the degenerate quantum numbers. We will also
find it convenient to use the SO(4) representation functions, the hyperspherical
harmonics with principal quantum number $n=J/2$. From
$L^2D^J_{KM}=J(J+1)D^J_{KM}$ we find
\begin{equation}
\kappa_n^2 = \frac{4J(J+1)}{a^2} = \frac{n(n+2)}{a^2},\quad
\lambda_n = \kappa_n^2 +\xi R = \frac{(n+1)^2}{a^2}\label{eq:s3-evalues}
\end{equation}
along with the spatial eigenmodes
\begin{equation}
u_{JKM} = \sqrt{\frac{n+1}{\Omega}} D^J_{KM}(\theta,\psi,\phi)
\end{equation}
for $S^3$.

To compute $\Xi_{nn'}$, we first use the sum rule
$\sum_{M''} D^J_{MM''}(D^J_{M'M''})^* = \delta_{M,M'}$ to get 
$\sum_{MK} D^J_{MK}(D^J_{MK})^* = 2J+1$. Since
\begin{eqnarray}
\sum_{\rm angular}\int d{\bf x} |u_{nKM} u_{n'K'M'}|^2 
&=& \frac{(n+1)(n'+1)}{\Omega}
\int d{\bf x} \sum_{KM}|D^J_{KM}|^2 \sum_{KM}|D^{J'}_{K'M'}|^2 \cr
&=& \frac{(n+1)^2 (n'+1)^2}{\Omega},
\end{eqnarray}
\ (\ref{eq:define-Xi}) is given by
\begin{equation}
\Xi_{nn'} = 
\frac{(n+1)^2 (n'+1)^2}{8\pi\Omega^2(uv)^\frac{1}{2}}
\left[  \frac{1}{uv} + \frac{n(n+2)}{a^2 v} + \frac{n'(n'+2)}{a^2 u}
+\frac{2n(n+2)n'(n'+2)}{a^4} \right].
\end{equation}

We now turn to the momentum correlation term $\Pi_{nn'}$. To facilitate
the evaluation of the spatial derivative terms, we use properties of the
generators of the Lie algebra, which for SO(3), is just the
quantum theory of angular momentum (see e.g. \cite{Zare}).
We have
\begin{eqnarray}
\nabla_i u_{nm} \nabla^i u^*_{n'm'} &=&
\gamma^{ab}{\bf e}_a(u_{nm}){\bf e}_b(u^*_{n'm'}) \cr\cr
&=& \left(\frac{2}{a}\right)^2 \frac{1}{\Omega} \sqrt{(n+1)(n'+1)} 
\sum_{a=1}^3 L_a(D^J_{KM}) L_a(D^{J'*}_{K'M'})
\end{eqnarray}
We also recast the spatial volume measure:
\begin{equation}
\int_{\cal M}\!d{\bf x}
=\frac{a^3}{4}\int_0^{2\pi}\!d\phi\int_0^\pi\!\sin\theta d\theta
\int_0^{2\pi}\! d\psi = \frac{a^3}{4}\int d{\bf\Omega},
\end{equation}
where we use the notation of \cite{Zare} for $d{\bf\Omega}$. We assume
the integrand is invariant under $\psi \rightarrow \psi + 2\pi$. This
condition is satisfied by $D^J_{KM}$.

Defining
\begin{equation}
B_{JJ'} = \sum_{K,M=-J}^J \sum_{K',M'=-J'}^{J'}
\int d{\bf\Omega}\left|
\sum_{a=1}^3 L_a(D^J_{KM})\; L_a(D^{J'*}_{K'M'})\right|^2
\label{eq-s3-Bjj-define}
\end{equation}
the momentum term \ (\ref{eq:define-Pi}) becomes
\begin{equation}
\Pi_{nn'} = \frac{(n+1)(n'+1)}{8\pi\Omega^2(uv)^\frac{1}{2}}
                                                   \frac{4}{a^4\pi^2}B_{JJ'}.
\end{equation}
Returning to \ (\ref{eq-s3-Bjj-define}), we can express 
it in terms of the angular momentum operators as
\begin{equation}
B_{JJ'} = \sum_{KMK'M'}\int d{\bf\Omega}\left| 
\frac{1}{2}L_+D^{J}_{KM} L_-D^{J'}_{K'M'}
+\frac{1}{2}L_-D^{J}_{KM} L_+D^{J'}_{K'M'}
+L_3D^{J}_{KM} L_3D^{J'}_{K'M'} \right|^2
\end{equation}
where $L_\pm = L_1 \pm iL_2$ are the raising and lowering operators.
Introducing the convenient notation $L_\alpha$,
$\alpha \in (+,-,0$) where $L_0$ replaces  $L_3$ and the symbols
\begin{equation}
C^\pm_K = \sqrt{J(J+1) - K(K\pm1)},\quad \quad C^0_K = K.
\end{equation}
the action of the angular momentum operators on the harmonic functions
is neatly given by
\begin{equation}
L_\alpha D^J_{KM} = (-i)^{-\alpha}C^{-\alpha}_K D^J_{K-\alpha,M}, \quad
L_\alpha D^{J*}_{KM} = -i^{-\alpha}C^{\alpha}_K D^J_{K+\alpha,M}.
\label{eq-s3-LaD}
\end{equation}
Introducing
\begin{equation}
<\alpha\beta ;\delta\epsilon> = \sum_{K,M=-J}^J \sum_{K',M'=-J'}^{J'}
\int d{\bf\Omega}
L_\alpha D^{J}_{KM} L_\beta D^{J*}_{KM}
L_\delta D^{J'}_{K'M'} L_\epsilon D^{J'*}_{K'M'},
\label{eq-s3-abde-define}
\end{equation}
we can express \ (\ref{eq-s3-Bjj-define}) as
\begin{eqnarray}
B_{JJ'} &=& <00;00> 
+\frac{1}{4}\left(<++;--> + <--;++> + <+-;+-> + <-+;-+> \right) \cr
&&+\frac{1}{2}\left(<0-;+0> + <0+;-0> + <+0;0-> + <-0;0+>\right).
\label{eq-s3-Bjj-1}
\end{eqnarray}
Using \ (\ref{eq-s3-LaD}), we can write \ (\ref{eq-s3-abde-define}) as
\begin{equation}
<\alpha\beta ;\delta\epsilon> = (-1)^{\beta+\epsilon}
i^{\alpha +\beta +\delta +\epsilon}
\sum_{KMK'M'}C^{-\alpha}_K C^{\beta}_K C^{-\delta}_{K'} C^{\epsilon}_{K'} 
\int d{\bf\Omega}
D^{J*}_{K+\beta,M} D^{J'*}_{K'+\epsilon,M'} 
D^{J}_{K-\alpha,M} D^{J'*}_{K'-\delta,M'}.
\end{equation}
The integral of the four-fold products of Wigner function is given by
$$
\int d{\bf\Omega} D^{J_1}_{K_1M_1} D^{J_2}_{K_2M_2} D^{J_3 *}_{K_3M_3} 
                                                            D^{J_4 *}_{K_4M_4}
$$
\begin{equation}
=8\pi^2 \sum_{JKM} (2J+1) \thj{J_1}{J_2}{J}{K_1}{K_2}{K}
\thj{J_1}{J_2}{J}{M_1}{M_2}{M}
\thj{J_3}{J_4}{J}{K_3}{K_4}{K}\thj{J_3}{J_4}{J}{M_3}{M_4}{M}.
\label{eq:fourfold-integral}
\end{equation}
This can readily be seen by two applications of the sum rule
\begin{equation}
D^{J_1}_{K_1M_1} D^{J_2}_{K_2M_2}
=\sum_{JKM}(2J+1)\thj{J_1}{J_2}{J}{K_1}{K_2}{K}
\thj{J_1}{J_2}{J}{M_1}{M_2}{M} D^{J *}_{KM}
\end{equation}
and the orthogonality property
\begin{equation}
\int d{\bf\Omega}  D^{J *}_{KM} D^{J'}_{K'M'} = \frac{8\pi^2}{2J+1}
\delta_{JJ'}\delta_{KK'}\delta_{MM'}.
\end{equation}
Utilizing this result, \ (\ref{eq-s3-abde-define}) has the form
\begin{equation}
<\alpha\beta ;\delta\epsilon> = (-1)^{\beta+\epsilon}
i^{\alpha +\beta +\delta +\epsilon} 8\pi^2
\sum_{K,M=-J}^J \sum_{K',M'=-J'}^{J'}C^{-\alpha}_K C^{\beta}_K C^{-\delta}_{K'} C^{\epsilon}_{K'} 
\end{equation}
$$
\times\sum_{J'',K'',M''} (2J''+1) 
\thj{J}{J'}{J''}{K+\beta}{K'+\epsilon}{K''}
\thj{J}{J'}{J''}{K-\alpha}{K'-\delta}{K''}
\thj{J}{J'}{J''}{M}{M'}{M''}
\thj{J}{J'}{J''}{M}{M'}{M''}.
$$
Using the orthogonality property of the 3-j symbols
\begin{equation}
\sum_{M,M'}\thj{J}{J'}{J_1}{M}{M'}{m_1}\thj{J}{J'}{J_2}{M}{M'}{m_2}
=(2J_1+1)^{-1}\delta_{J_1,J_2}\delta_{m_1,m_2},
\end{equation}
we can reduce the above four-fold product of 3-j symbols to a two-fold
product by doing the $M,M',M''$ sums to find the final form that is
most useful to us:
\begin{equation}
<\alpha\beta ;\delta\epsilon> = (-1)^{\beta+\epsilon}
8\pi^2 \sum_{J^{''}} (2J''+1) 
\label{eq-s3-abde-final}
\end{equation}
\[
\times\sum_{KK'K''}C^{-\alpha}_K C^{\beta}_K C^{-\delta}_{K'} C^{\epsilon}_{K'} 
\thj{J}{J'}{J''}{K+\beta}{K'+\epsilon}{K''}
\thj{J}{J'}{J''}{K-\alpha}{K'-\delta}{K''}.
\]
The triangularity of the 3-j symbols implies the condition
$\alpha+\beta+\delta+\epsilon=0$ for $<\alpha\beta ;\delta\epsilon>$
not to vanish.

With this relation, we can evaluate the terms we need for 
(\ref{eq-s3-Bjj-1}):
\begin{mathletters}
\label{eq:Bjj-terms}
\begin{eqnarray}
<00;00> &=& 8\pi^2 \sum_{J''} (2J''+1)\sum_{KK'K''} 
\left[K K'\thj{J}{J'}{J''}{K}{K'}{K''}\right]^2\\
<++;--> &=& 8\pi^2 \sum_{J''} (2J''+1)\sum_{KK'K''} 
C^-_K C^+_K C^+_{K'} C^-_{K'}
\thj{J}{J'}{J''}{K+1}{K'-1}{K''}
\thj{J}{J'}{J''}{K-1}{K'+1}{K''}\\
        &=& <++;--> \\
<+-;+-> &=& 8\pi^2 \sum_{J''} (2J''+1)\sum_{KK'K''} 
\left[C^-_K C^-_{K'}
\thj{J}{J'}{J''}{K-1}{K'-1}{K''}\right]^2\\
<-+;-+> &=& 8\pi^2 \sum_{J''} (2J''+1)\sum_{KK'K''} 
\left[C^+_K C^+_{K'}
\thj{J}{J'}{J''}{K+1}{K'+1}{K''}\right]^2\\
<0-;+0> &=& 8\pi^2 \sum_{J''} (2J''+1)\sum_{KK'K''} 
KK' C^-_K C^-{K'} \thj{J}{J'}{J''}{K-1}{K'}{K''}
\thj{J}{J'}{J''}{K}{K'-1}{K''}\\
        &=& <+0;0->\\
<0+;-0> &=& 8\pi^2 \sum_{J''} (2J''+1)\sum_{KK'K''} 
KK' C^+_K C^+{K'} \thj{J}{J'}{J''}{K+1}{K'}{K''}
\thj{J}{J'}{J''}{K}{K'+1}{K''}\\
        &=& <+0;0->
\end{eqnarray}
\end{mathletters}

We expect $B_{JJ'}$ to assume the form
\begin{equation}
B_{JJ'} = 8\pi^2 (aJ^3 + bJ^2 + cJ +d)(aJ'^3 + bJ'^2 + cJ' +d).
\end{equation}
Determining the coefficients by evaluating $B_{JJ'}$ for four different
pairs of $J,J'$, we find
\begin{equation}
B_{JJ'} = 8\pi^2 J(J+1)(J+2)J'(J'+1)(J'+2)
= \frac{\pi^2}{2} n(n+1)(n+2) n'(n'+1)(n'+2)
\end{equation}
and
\begin{equation}
\Pi_{nn'} = \frac{(n+1)^2 (n'+1)^2}{8\pi\Omega^2(uv)^\frac{1}{2}}
\frac{2}{a^4}n(n+2) n'(n'+2),
\end{equation}
along with
\begin{equation}
\Psi_{nn'} = 
\frac{(n+1)^2 (n'+1)^2}{8\pi\Omega^2(uv)^\frac{1}{2}}
\left[  \frac{1}{uv} + \frac{n(n+2)}{a^2 v} + \frac{n'(n'+2)}{a^2 u}
+\frac{4n(n+2)n'(n'+2)}{a^4} \right].
\end{equation}
Combining these results, the needed zeta function is
\begin{eqnarray}
\zeta_\Psi(0,\nu) &=& \frac{1}{8\pi\Omega^2 a^{2-4\nu}}
\sum_{n,n'=1}^\infty  n^2 n'^2 \int_0^\infty\!\!du\int_0^\infty\!\!dv
(uv)^{\nu-\frac{3}{2}} \cr
&& \times\left(1 + (n^2-1)u + (n'^2-1)v + 4(n^2-1)(n'^2-1)uv\right)
e^{-n^2 u - n'^2 v}.
\end{eqnarray}
Using relation \ (\ref{eq:Riemann-zeta}), the analytic continuation takes the form
\[
\zeta_\Psi(0,\nu) = 
\frac{1}{8\pi\Omega^2 a^{2-4\nu}}\left\{
\Gamma\left(\nu-\frac{1}{2} \right)^2\zeta_R(2\nu-3)^2 
\right.
\]
\begin{equation}   
\left.
+2 \Gamma\left(\nu-\frac{1}{2} \right)\Gamma\left(\nu+\frac{1}{2} \right)
     \zeta_R(2\nu-3) \left(\zeta_R(2\nu-3)-\zeta_R(2\nu-1)\right)
+4 \Gamma\left(\nu+\frac{1}{2} \right)^2 \left(\zeta_R(2\nu-3) 
  - \zeta_R(2\nu-1)\right)^2
\right\}
\end{equation}
Setting $\nu=0$, we get finally the variance of the energy density of a scalar 
field on the Einstein Universe:
\begin{equation}
\Delta\rho^2 = \frac{37}{76800 \pi^4 a^8}.
\end{equation}
Using the result \cite{Ford76} for the energy density,
\begin{equation}
\rho = \frac{1}{480 \pi^2 a^4}
\end{equation}
 we find the dimensionless measure of the fluctuations in the
energy density \ (\ref{eq:define-Delta'}) is given by
\begin{equation}
\Delta'(\Sigma=S^3) = 111.
\end{equation}
Thus for the Einstein Universe, the fluctuations in the energy density are
indeed quite large. Effects due to the fluctuations of the metric will become
important {\it before} the size of the Universe approaches that of the Planck
scale.

\section{Discussion}

In this paper we have shown how the correlation function for the 
 quantum stress energy
tensor is related to the second metric variation of the effective action.
This parallels the definition of the expectation value of the quantum
stress energy as the first metric variation of the effective action.
A physically meaningful expectation value is derived from a regularized 
or renormalized effective action.
Likewise, the correlation function is defined here in terms of the
second variation of the regularized effective action. 
The correlation of the stress energy tensor  computed for two
distinct points is  finite regardless of whether the effective action 
is regularized or not (excluding lightlike seperated points for a massless
quantum field). It is only when the autocorrelation is computed that
the issue of regularization arises. Nonetheless, we choose to use the
correlation function defined in terms of the regularized effective action.
This is a consistent approach since then it is defined as the second
variation of the same object for which the expectation value is the first
variation.

Since we have only considered geometries for which a Euclidean section
exists, we can regularize the effective action via the zeta function
method. For a quantum system its generalized zeta function is given
as the Mellin transform of the trace of its heat kernel. The effective
action is then given by the derivative of the zeta function. This constituted 
our starting point for computing the second variation. The key to
the zeta function method is to control the divergence of the heat kernel
present when the Schwinger proper time vanishes. This is done by introducing
positive powers of the proper time and then performing an analytic continuation
in powers of the proper time. At the end of the calculation, the
variable of analytic continuation is set to zero. This is consistent in that
the introduced power can be relaxed to zero at any point of the calculation
to recover the initial formal expression.

The second variation of the generalized zeta function is facilitated by 
the Schwinger perturbative expansion, which shows how the trace of the 
heat kernel responses. Once we have this response, we only need the
trace of a pair of heat kernels. To make this resultant expression
meaningful, the key idea of the 
zeta function is: for each of these traces 
we introduce a power of the proper time variable for that trace. 
The stress energy correlation function is expressed in
terms of traces of the system's heat kernel, regularized via the generalized
zeta function method. This is one
of the main results of this paper.
The zeta function method allows us to
relax the introduced powers at any time and recover the formal expression
for the correlation function.

For geometries admitting a mode decomposition of the invariant operator,
we display the correlation function explicitly in terms of these modes.
The specialization to homogeneous geometries is considered and the
simplification this entails is explored. 

Resent results \cite{KuoFord,Barton} have suggested that quantum fluctuations
of the energy density may be significant for systems with non-zero vacuum
energy density. Our result confirms this assertion and goes beyond.
 In particular, the
variance of the energy density for a massless scalar field is found.
(The variance is the coincidence limit of the energy-energy correlation
function.) This measure of the quantum fluctuations 
is calculable since we have developed the correlation function in terms of the
zeta function regularized effective action. Our results are in excellent
agreement with the results of \cite{KuoFord} for Minkowski spacetime
with one compact dimension. We have extended this work to
flat spacetimes of arbitrary dimension with one periodic dimension and 
found that the variance grows quadratically with the dimension of spacetime.
 This may have unexpected implications for Kaluza-Klein theory. 
We also found the fluctuations for the Einstein Universe, which turns out
to be more than ten times larger than the energy density. This shows 
that quantum fluctuations will become important
at energy scales below the Planck scale and supports the suggestion
\cite{HuBanff} that critical dynamics at such scales could reveal interesting
new phenomena.
Knowledge of the higher order correlation functions of the
quantum stress energy may be necessary
to account for the full backreaction effect
of these large fluctuations of the quantum fields on the dynamics
of the geometry (see, e.g., \cite{Erice})
and for investigating the issue of the viability of
semiclassical theories \cite{HuTsukuba,CH94} at the Planck scale.\\

\noindent{\bf Acknowledgement} NP thanks Professor Elizalde
for correspondence on the zeta function method.  Part of the 
work 
was done when BLH was on leave at the Institute for Advanced 
Study, Princeton.
This work is
supported in part by NSF grant PHY94-21849.

\newpage

\setcounter{equation}{0}
\renewcommand{\theequation}{A\arabic{equation}}

\appendix

\section{Calculation of $\int d\Omega_d\int d\Omega'_d\cos^2\gamma$}

In this appendix, we calculate $\beta$ defined by
\begin{equation}
B_d = \int_{S^d} d\Omega_d \int_{S^d} d\Omega'_d \cos^2\gamma = 
\beta {\rm Vol}^2(S^d)
\end{equation}
where $d\Omega_d$ is the volume measure on the $d$ sphere $S^d$ and 
$\cos\gamma = |\vec x\!\cdot\!\vec x'|$. Here $\vec x$ and $\vec x'$ are
unit vectors in $R^{d+1}$(i.e. they are `points' on $S^d$). 
If we parameterize $S^d$ with the Euler
angles $\theta_i$, $i=1,\ldots\,d$ with $0\le\theta_1 \le 2\pi$ and
$0\le\theta_j \le \pi$, $j \ne 1$, we have
\begin{equation}
d\Omega_d = \sin^{d-1}\theta_d\cdots\sin\theta_2 d\theta_d\cdots d\theta_1.
\end{equation}
Using
\begin{equation}
\int_0^\pi \sin^{k-1}\theta d\theta = 
\frac{\sqrt{\pi}\Gamma\left(\frac{k}{2} \right)}{\Gamma\left(\frac{k+1}{2}\right)}
\label{eq:beta-sin^k}
\end{equation}
we get
\begin{equation}
{\rm Vol}(S^d) = \int_{S^d} d\Omega_d = 
\frac{2\pi^{(d+1)/2}}{\Gamma\left(\frac{d+1}{2} \right)}.
\end{equation}

Let $\vec x'$ be a point on $S^d$ and define 
$\alpha(\vec x') = \int d\Omega_d |\vec x\!\cdot\!\vec x'|^2$. We have
$\alpha(\vec x')$ is independent of $\vec x'$ and hence
\begin{equation}
B_d = \int_{S^d}d\Omega'_d \alpha(\vec x')
= {\rm Vol}(S^d)\alpha(\vec x_0), \quad {\rm any}\  \vec x_0\in S^d.
\end{equation}
We take $\vec x_0 = (0,\ldots,x_{0,d+1}=1)$. Then 
$|\vec x\!\cdot\!\vec x'|^2 = \cos^2\theta_d=1-\sin^2\theta_d$
and
\begin{eqnarray}
\alpha(\vec x_0)&=&\int_{S^d}d\Omega_d - \int_{S^d}d\Omega_d \sin^2\theta_d
\nonumber\\
&=&{\rm Vol}(S^d)\left[1 -
\frac{\int_0^\pi \sin^{d+1}\theta_d d\theta_d} 
{\int_0^\pi \sin^{d-1}\theta_d d\theta_d} \right]\nonumber\\
&=&{\rm Vol}(S^d)\left[1 -
\frac{\Gamma\left(\frac{d+2}{2} \right)}{\Gamma\left(\frac{d}{2} \right)}
\frac{\Gamma\left(\frac{d+1}{2} \right)}{\Gamma\left(\frac{d+3}{2} 
                                                    \right)}\right]\nonumber\\
&=&{\rm Vol}(S^d)\frac{1}{d+1}
\end{eqnarray}
Thus we get
\begin{equation}
\beta = \frac{1}{d+1}
\label{eq:beta}
\end{equation}

\end{document}